\documentclass[prb,twocolumn,amsmath,amssymb,superscriptaddress]{revtex4-1}
\usepackage{graphicx}% Include figure files
\usepackage{dcolumn}% Align table columns on decimal point
\usepackage{hhline}
\usepackage{bm,color}% bold math

\usepackage{float}

\usepackage{amsmath}

\usepackage{amsfonts}

\usepackage{amssymb}

\usepackage{epstopdf}

\begin{document}

%\wideabs{

\title{Quantum criticality in the coupled two-leg spin ladder Ba$_2$CuTeO$_6$}

\author{A. Glamazda}
\affiliation{Dept. of Physics, Chung-Ang University, Seoul 156-756, Republic of Korea}

\author{Y. S. Choi}
\affiliation{Dept. of Physics, Chung-Ang University,  Seoul 156-756, Republic of Korea}

\author{S.-H. Do}
\affiliation{Dept. of Physics, Chung-Ang University,  Seoul 156-756, Republic of Korea}

\author{S. Lee}
\affiliation{Dept. of Physics, Chung-Ang University,  Seoul 156-756, Republic of Korea}

\author{P. Lemmens}
\affiliation{Inst. for Condensed Matter Physics, TU Braunschweig, D-38106 Braunschweig, Germany}
\affiliation{Laboratory for Emerging Nanometrology (LENA), TU Braunschweig, 38106 Braunschweig, Germany}

\author{A. N. Ponomaryov}
\affiliation{Dresden High Magnetic Field Lab. (HLD-EMFL), Helmholtz-Zentrum Dresden-Rossendorf, Dresden D-01328, Germany}

\author{S. A. Zvyagin}
\affiliation{Dresden High Magnetic Field Lab. (HLD-EMFL), Helmholtz-Zentrum Dresden-Rossendorf, Dresden D-01328, Germany}

\author{J. Wosnitza}
\affiliation{Dresden High Magnetic Field Lab. (HLD-EMFL), Helmholtz-Zentrum Dresden-Rossendorf, Dresden D-01328, Germany}

\author{Dita Puspita Sari}
\affiliation{Department of Physics, Graduate School of Science, Osaka University, Toyonaka, Osaka 560-0043, Japan}
\affiliation{Advanced Meson Science Laboratory, RIKEN, 2-1 Hirosawa, Wako, Saitama 351-0198, Japan}

\author{I. Watanabe}
\affiliation{Advanced Meson Science Laboratory, RIKEN, 2-1 Hirosawa, Wako, Saitama 351-0198, Japan}

\author{K.-Y. Choi}
\email[]{kchoi@cau.ac.kr}
\affiliation{Dept. of Physics, Chung-Ang University,  Seoul 156-756, Republic of Korea}

%\date{\today}

\begin{abstract}
We report on zero-field muon spin rotation, electron spin resonance and polarized Raman scattering measurements of the coupled quantum spin ladder Ba$_2$CuTeO$_6$.  Zero-field muon spin rotation and electron spin resonance probes disclose a successive crossover from a paramagnetic through a spin-liquid-like into a magnetically ordered state with decreasing temperature. More significantly, the two-magnon Raman response obeys a $T$-linear scaling relation in its peak energy, linewidth and intensity.  This critical scaling behavior presents an experimental signature of proximity to a quantum critical point from an ordered side in Ba$_2$CuTeO$_6$.
\end{abstract}

\maketitle

\section{Introduction}
Quantum criticality and quantum phase transitions dictated by quantum mechanical fluctuations are the key notions of current condensed matter physics.  Quantum critical systems exhibit a universal scaling behavior, which results
from the intertwined effects of thermal and quantum fluctuations~\cite{Sondhi}. In
a quantum critical regime, the characteristic energy scale of low-energy excitations is determined solely by temperature.  Such scaling relations are exemplified in distinct strongly correlated systems including high-temperature superconductors, heavy-fermion metals, and quantum magnets~\cite{Lohneysen,Gegenwart,Sachdev,Lake,Merchant,Keimer,Helton,Aronson}.

Quantum spin ladders consisting of leg ($J_l$) and rung ($J_r$) couplings offer an outstanding platform to
study quantum-critical spin dynamics and have a far-reaching relevance to diverse fields of physics such as Tomonaga-Luttinger liquids, magnon fractionalization, unconventional superconductivity, quantum computing, and quark confinement
\cite{Maekawa,Uehara,Li,Lake09,Thielemann,Klanjsek,Hong,Schmidiger,Jeong,Choi}.
Isolated two-leg ladders with $J_l/J_r\neq 0$ are known to have a short-range resonating valence bond (RVB) state. Depending on the $J_l/J_r$ ratio, their elementary excitations are given by either triplons or pairs of bound spinons~\cite{White}. It has been proposed that a quantum phase transition occurs from the RVB  to the magnetically ordered state with growing interladder couplings~\cite{Normand}. In the ordered phase, the low-energy excitations are gapless spin waves arising from  spontaneous symmetry breaking.  In the vicinity of a quantum critical point, a number of thermally excited magnons increase progressively with temperature
and thus their interaction energy becomes comparable to the energy of a single magnon, acquiring the quantum nature of the quasiparticle excitations. Experimentally, such a quantum-critical state has so far remained elusive in coupled two-leg ladders due to the lack of relevant materials.

\begin{figure*}
\label{figure1}
\centering
\includegraphics[width=12cm]{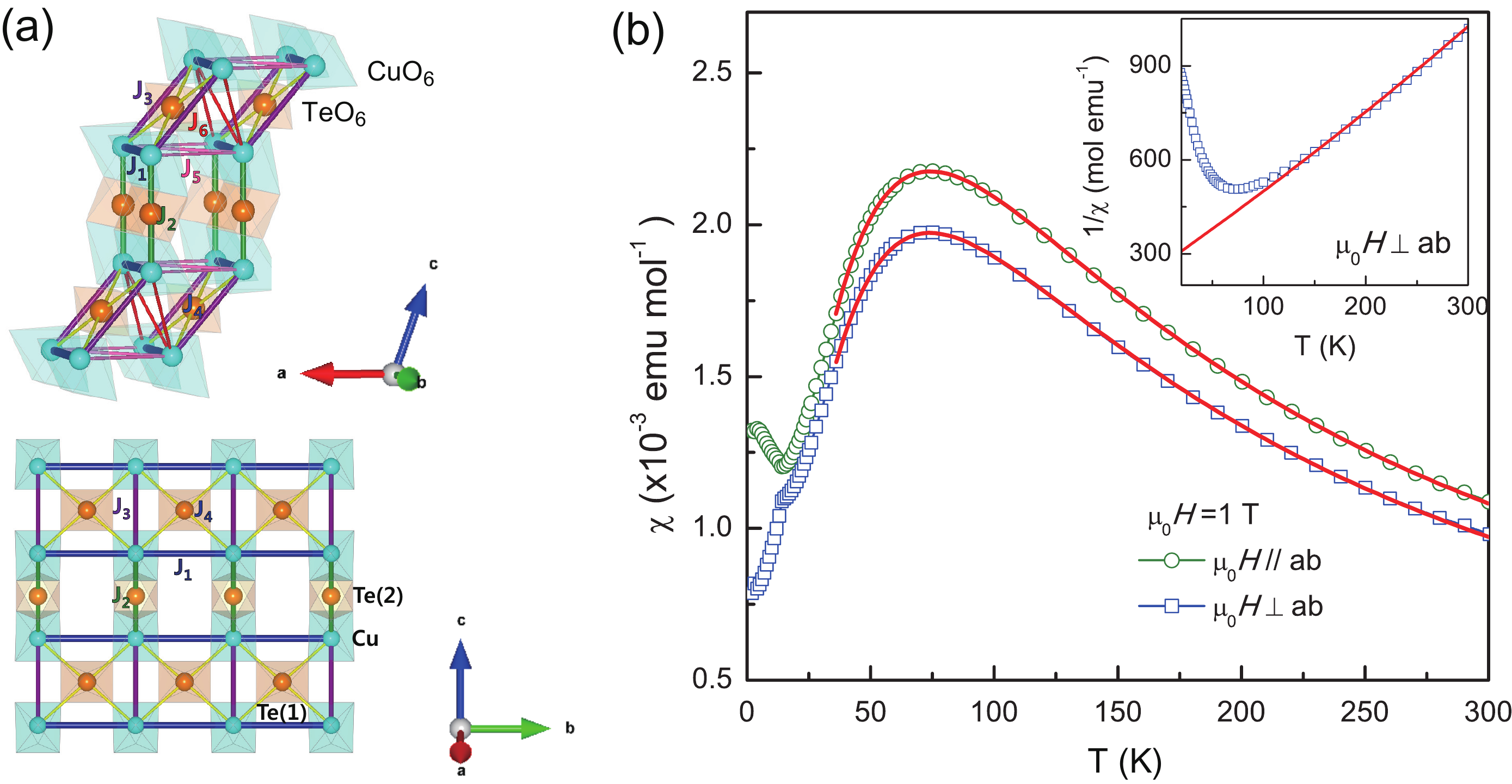}
\caption{(a) (Upper panel) Crystal structure of Ba$_2$CuTeO$_6$ viewed along the $b$ axis with the main six exchange constants $J_i$ ($i=1-6$) and (Lower panel) schematic sketch of the two-leg spin ladder in the $bc$ plane. CuO$_6$ (cyan) and TeO$_6$ (maroon) octahedra are stacked in layers along the $c$ axis. (b) Temperature dependence of the magnetic susceptibility $\chi(T)$ for Ba$_2$CuTeO$_6$ measured in an external field of $\mu_0H =1$~T applied parallel and perpendicular to the $ab$ plane. The red solid lines are fits using a two-leg ladder model. The inset shows the inverse susceptibility with a fitting to a Curie-Weiss law.}
\end{figure*}

Ba$_2$CuTeO$_6$ appears to be a prime candidate for a three-dimensionally networked spin ladder.
In this material, two-leg spin ladders are formed by the tellurium-bridged Cu$^{2+}$($S=1/2$) ions~\cite{Iwanaga,Gibbs,Rao}. The TeO$_6$ octahedral coordination is known for its capability to build a rich spin connectivity, while Jahn-Teller active CuO$_6$ octahedra favor a low lattice coordination~\cite{Khomskii}. This structural peculiarity leads to a number of the residual interladder exchange interactions, thereby placing Ba$_2$CuTeO$_6$ close to a quantum critical point.

In the monoclinic phase of Ba$_2$CuTeO$_6$, triangular arrangements of the CuO$_6$ and the TeO$_6$ octahedra are stacked alternately along the $c$ axis~\cite{Kohl} [see Fig.~1(a)]. The $S=1/2$ Cu$^{2+}$ ions are coupled through complex Cu-O-Te-O-Cu superexchange paths, giving rise to six main exchange interactions  $J_i$ ($i=1-6$). Depending on the relative strength of $J_2$ and $J_3$, two different ladder models have been proposed: (i)  a two-leg spin ladder with $J_1=J_l$ and $J_3=J_r$~\cite{Gibbs} and (ii) a two-leg spin ladder with $J_1=J_l$ and $J_2=J_r$  within the $bc$ plane~\cite{Rao}. The residual in- and inter-plane interladder couplings stabilize a possible low-temperature magnetic ordering.

The spin-ladder correlations are evidenced by the observation of a broad peak at $T_{\mathrm{max}}=75$~K in the magnetic susceptibility and the spin-gap excitation of $\Delta=50$~K inferred from the $^{125}$Te spin-lattice relaxation rate $1/T_1$ at elevated temperatures~\cite{Gibbs}. At low temperatures, the RVB state is preempted by the development of long-range magnetic ordering at $T_\mathrm{N}=15-16$~K, indicated by a small kink  in the magnetic susceptibility~\cite{Gibbs,Rao}. However, the magnetic transition is largely hidden, while showing no magnetic Bragg peaks, no divergence of $1/T_1$, and no apparent $\lambda$-like anomaly in the specific heat. This marginal transition points towards a thermally driven dimensional crossover from a spin ladder to a 3D magnetism with strong quantum fluctuations.

Here, we present the results of zero-field (ZF) muon spin rotation ($\mu$SR), electron spin resonance (ESR), and inelastic light scattering measurements of the quantum spin ladder Ba$_2$CuTeO$_6$. By combining two resonance techniques, we identify a consecutive evolution of magnetic correlations from a paramagnetic through a spin ladder-like to a magnetically ordered state. The most salient finding is that the two-magnon Raman response
exhibits a $T$-linear dependence in its peak energy, linewidth, and intensity  over almost two decades of temperature. This scaling behavior suggests that Ba$_2$CuTeO$_6$ lies close to a quantum critical point from an ordered side.

\section{\label{sec:level2}Experimental Details}
Powder samples of Ba$_2$CuTeO$_6$ were synthesized by the solid-state reaction method. Stoichiometric mixtures of BaCO$_3$, CuO, and TeO$_2$ powders were heated to 1000$^{\circ}$ for 24 hours with several intermediate grindings under flowing oxygen. Single crystals were grown by the BaCl$_2$ flux method. The prepared polycrystalline samples of Ba$_2$CuTeO$_6$ were mixed with the flux of BaCl$_2$ in the molar ratio of 1:10 and then were melted in an alumina crucible. The mixture was heated at 1200$^{\circ}$ for 5 hours and then slowly cooled down to 900$^{\circ}$ at a rate of 2$^{\circ}$/h. Dark green crystals were obtained and were separated from the flux with hot water.  Powder x-ray diffraction and magnetic susceptibility confirmed a high quality of the grown samples.

X-band ESR experiments were performed using a Bruker Elexsys E500 spectrometer  at the Helmholtz-Zentrum Dresden-Rossendorf.
The spectrometer measures the field derivative of the absorbed microwave power, $dP(H)/dH$. In the measured temperature range, the ESR signals are  well described by a
Lorentzian line profile $P(H)=-\frac{16Ah}{(3+h^2)^2}$, where $h=\frac{2(H-H_{\mathrm{res}})}{H_{\mathrm{res}}}$.
Here, $A$, $H_{\mathrm{res}}$, and $\Delta H_{pp}$ are the amplitude, the resonance field, and the peak-to-peak linewidth, respectively.
The $g$-factors are determined by the relation $g = h\nu/\mu_BH_{\mathrm{res}}$ with $\nu=9.4$~GHz. Since $\Delta H_{pp}$ is inversely proportional to
the spin-spin relaxation rate $T_2^*$, it gives information about spin-spin correlations.

ZF-$\mu$SR measurements were carried out on the ARGUS spectrometer of RIKEN-RAL at the
Rutherford Appleton Laboratory. In a $\mu$SR experiment spin-polarized positive muons (momentum 28 MeV/c) are implanted into a sample.
The experimentally measured quantity is the muon spin polarization function  $a_0P(t)= \frac{N_\mathrm{B}(t)-\alpha N_\mathrm{F}(t)}{N_\mathrm{B}(t)+\alpha N_\mathrm{F}(t)}$,
where $a_0$ is the initial asymmetry, $\alpha$ is an efficiency ratio of the forward and the backward detectors, and $N_\mathrm{F}$(t) and $N_\mathrm{B}$(t) are the muon counts at the detectors antiparallel and parallel to an incident muon spin direction. $a_0P(t)$ contains information on the magnitude, static distribution, and fluctuations of local magnetic fields.
The collected data were analyzed using the free software package WiMDA~\cite{Pratt}.

A polarization-resolved Raman spectroscopy was employed to probe spin and phonon excitations of single crystals of Ba$_2$CuTeO$_6$. Raman scattering experiments were carried out with the excitation line $\lambda=532.1$~nm of a Nd:YAG (neodymium-doped yttrium aluminium garnet) solid-state laser in a quasi-backscattering geometry.
The scattered spectra were collected using a DILOR-XY triple spectrometer  equipped with a liquid-nitrogen-cooled CCD.  The samples were mounted onto an evacuated closed-cycle cryostat, while varying a temperature between  9 and 293~K. To minimize laser heating effect, we
used the laser power of $P=1.5$~mW, focusing to a 0.1-mm-diameter spot on the surface of the single crystal. The heating of the sample did not exceed ~2 K.

\section{Results and discussion}

\subsection{Static magnetic susceptibility }
Figure~1(b) shows the temperature dependence of the magnetic susceptibility $\chi(T)$ measured in an external field of $\mu_0H =1$~T applied parallel and perpendicular to the $ab$ plane. Our data are in a good accordance with the previously reported results~\cite{Gibbs,Rao}. With decreasing temperature, $\chi(T)$ first displays a broad maximum at about 75 K with the subsequent drop and then a small kink at about $T_\mathrm{N}\sim 15$~K. The 75 K broad maximum is a common feature of low-dimensional antiferromagnets developing short-range spin correlations. The bifurcation of $\chi(T)$ between the two orientations for $T< 15$~K together with the kink is associated with the onset of long-range magnetic ordering.

The high-temperature part of $\chi_{\bot ab}(T)$ above 170 K can be well described by a Curie-Weiss law, yielding the Curie-Weiss temperature $\Theta_{\mathrm{CW}}\approx -118$~K and the effective magnetic moment $\mu_{\mathrm{eff}}=1.94\, \mu_{\mathrm{B}}/\mathrm{Cu}^{2+}$. We analyze the magnetic susceptibility using a two-leg ladder model derived from a quantum Monte Carlo method~\cite{Johnston}. After the correction for the core diamagnetism $\chi_{\mathrm{dia}}=-1.48\times 10^{-4}~ \mathrm{emu}\cdot \mathrm{mol}^{-1}$, the ladder model fit to $\chi_{\bot ab}(T)$ gives $g\approx 2.13$, the leg coupling $J_{\mathrm{leg}}/k_\mathrm{B}\approx 93$~K, $J_{\mathrm{rung}}/J_{\mathrm{leg}}\approx0.97$, and the spin gap $\Delta/k_{\mathrm{B}}\approx 46$~K. The fitting parameters for $\chi_{\| ab}(T)$ are determined to be $g\approx 2.23$, the leg coupling $J_{\mathrm{leg}}/k_\mathrm{B}\approx 94$~K, $J_{\mathrm{rung}}/J_{\mathrm{leg}}\approx0.94$, and $\Delta/k_{\mathrm{B}}\approx 47$~K. The exchange coupling constant agrees perfectly with the value of $J/k_{\mathrm{B}}=98$~K estimated by a two-magnon excitation
(see the section III.D).

\subsection{Muon spin rotation/relaxation}
\begin{figure}
\label{figure1}
\centering
\includegraphics[width=8.5cm]{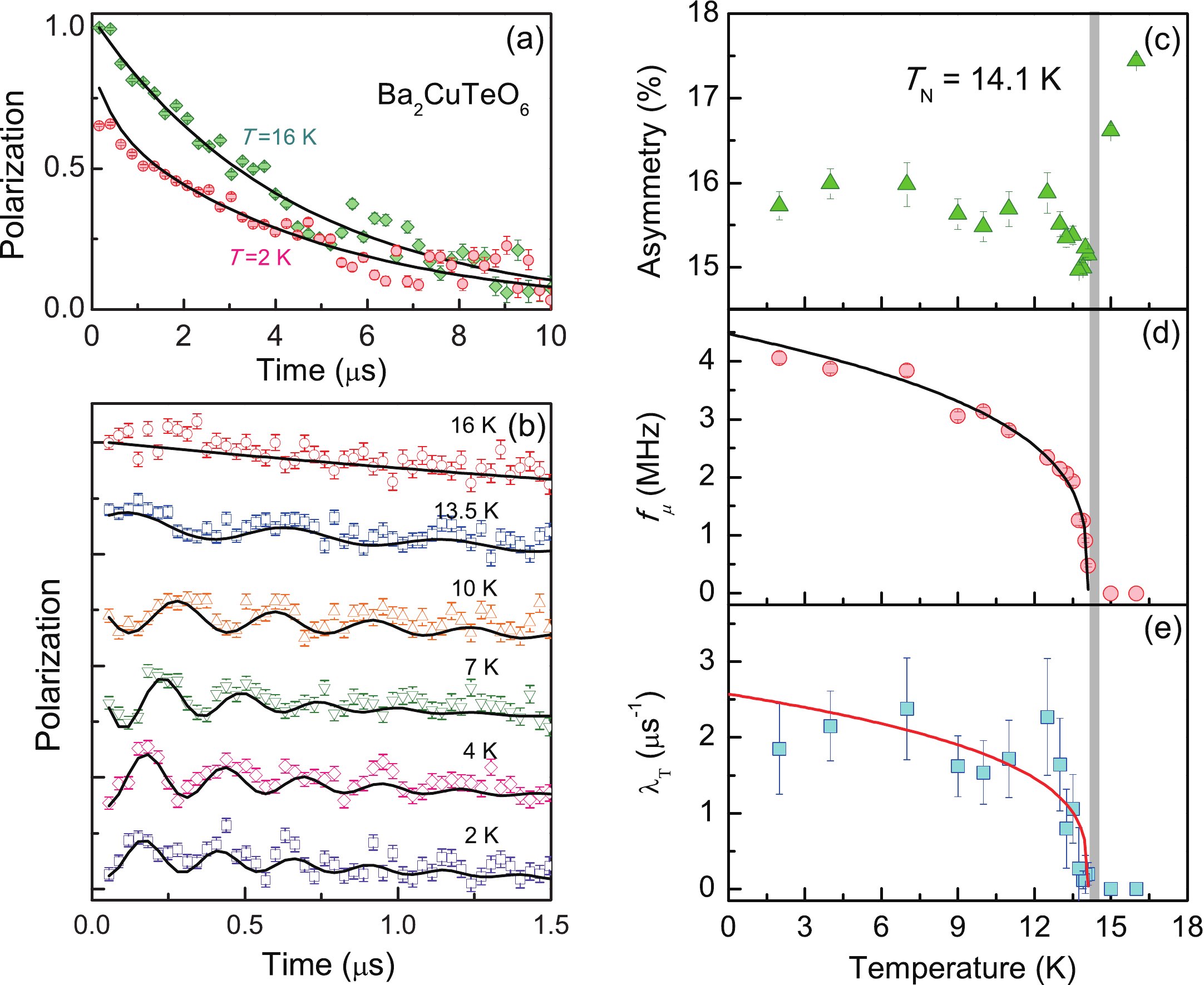}
\caption{(a) Representative data of the muon polarization of Ba$_2$CuTeO$_6$ measured above and below $T_\mathrm{N}$. The solid lines are fits described in the main text. (b) Early-time behavior of the muon polarization at various temperatures. The spectra are vertically shifted. (c),(d),(e) The asymmetry,  muon spin precession frequency $f_\mu$, and transverse relaxation rate $\lambda_T$ as a function of temperature.  The vertical bar indicates the onset of magnetic ordering at $T_\mathrm{N}=14.1$~K. The solid lines are fits discussed in the text.}
\end{figure}
To ensure the occurrence of static magnetic order, local magnetic fields were detected by ZF muon-spin rotation on a powder sample of Ba$_2$CuTeO$_6$.
The time decay of the muon spin polarization $P(t)$  at temperatures above and below $T_\mathrm{N}$ is shown in Fig.~2(a). Upon cooling towards $T_\mathrm{N}$, we observe spontaneous muon-spin precession in  $P(t)$ together with a drop in the early-time asymmetry as shown in Fig.~2(b). This confirms the development of static local magnetic fields at the muon stopping site. The polarization curves can be well described by the sum of an exponentially relaxing cosine function and a simple exponential function:
$P(t) = a_L\exp(-\lambda_L t)+ a_T\exp(-\lambda_T t)\cos(2\pi f_\mu t+\phi)$, where the two terms represent muons polarized transverse and parallel to the local magnetic fields. Here, $a_T$ ($a_L$) and $\lambda_{T}$ ($\lambda_{L}$) are the transverse (longitudinal) relaxing fraction and the transverse (longitudinal) relaxation rate caused by slow dynamics of the magnetic moments, respectively. $f_\mu$ is the muon-spin precession frequency.

The temperature dependences of the asymmetry, $f_\mu$, and $\lambda_T$  are plotted in Figs.~2(c)-(e). All $\mu$SR parameters show distinct changes at $T_\mathrm{N}$.  The initial asymmetry drops rapidly on cooling to $T_\mathrm{N}$. The missing asymmetry is ascribed to an unresolved precession signal within the pulsed muon beam time window.
$f_\mu(T)$, corresponding to the magnetic order parameter, is fitted to the phenomenological form $f_\mu(T)=f_0 (1-(T/T_\mathrm{N}))^\beta$,
where \textit{f}$_0=4.3$~MHz is the frequency at $T=0$~K and $\beta = 0.29(1)$ is the critical exponent.
The extracted value of $\beta$  hardly varies with the
choice of a temperature range  (not shown here). We further note that the obtained critical exponent is not much different from the value $\beta=0.365$ of the 3D Heisenberg model. $T_\mathrm{N}=14.1$~K is slightly lower than the transition temperature of 15~K determined from the uniform susceptibility. The temperature dependence of $\lambda_{T}(T)$ can be also modeled with the same order-parameter fit as plotted in Fig.~2(e).

\subsection{Electron spin resonance}

\begin{figure}
\label{figure1}
\centering
\includegraphics[width=8cm]{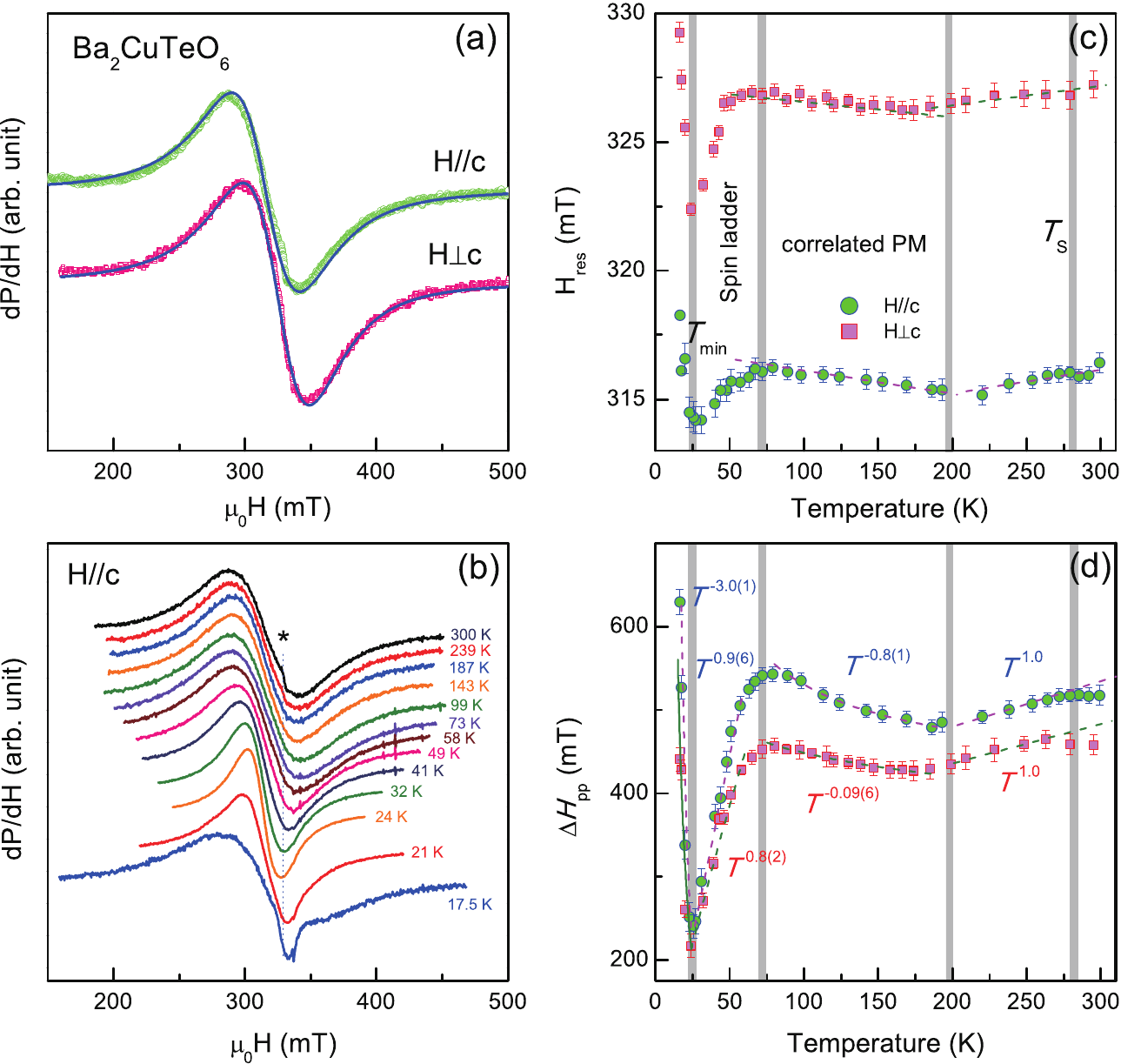}
\caption{(a) Derivative of the ESR absorption spectra, $dP/dH$ at room temperature for  $H\parallel c$ and $H \perp c$. The solid lines are fits using a Lorentzian profile.
(b) $dP/dH$ spectra for different temperatures and $H\parallel c$. The spectra are vertically shifted for clarity. The asterisk denotes an impurity signal. (c) Temperature dependence of
the resonance field, $H_{\mathrm{res}}(T)$, for $H\parallel c$ and $H\perp c$.
(d) Temperature dependence of the peak-to-peak linewidth, $\Delta H_{pp}(T)$.
The dashed lines are fits using  $\Delta H_{pp}(T)\sim T^{\alpha}$. The shaded bars represent a crossover of magnetic correlations.}
\end{figure}

Further information on the evolution of spin and structural correlations is provided by X-band ESR measurements. In Fig.~3(a), we compare the room-temperature ESR spectra for $H\parallel c$ and $H \perp c$, which show a single Lorentzian line shape due to fast electronic fluctuations
of Cu$^{2+}$ spins induced by the exchange interactions. At room temperature, the $g$-factors are
evaluated to be $g_c =2.087 \pm 0.003$ and $g_{ab}=2.134\pm 0.005$, typical for Cu$^{2+}$ ions with a quenched orbital moment~\cite{AB}. As evident from Fig.~3(b), upon cooling the ESR signal evolves in a nonmonotonic manner. In addition to the main signal, we observe the weak, narrow peak at $H_{\mathrm{res}}=329$~mT (denoted by the asterisk). The extra peak grows systematically in intensity with decreasing temperature without changing $H_{\mathrm{res}}$. Thus, it is ascribed to a small concentration of defects and orphan spins, to which ESR is sensitive.

For detailed quantification, the ESR spectra were fitted to the derivative of Lorentzian profiles. The resulting temperature dependencies of the resonance field, $H_{\mathrm{res}}$ and the peak-to-peak linewidth, $\Delta H_{pp}$ are shown in Figs.~3(c),(d).
Strikingly, $H_{\mathrm{res}}(T)$ and $\Delta H_{pp}(T)$ shows similar temperature dependences, indicating the intriguing evolution of spin correlations as the resonance field is shifted by the buildup of internal fields. Even at temperatures above 200~K, both $H_{\mathrm{res}}(T)$ and $\Delta H_{pp}(T)$ exhibit a linear increase with increasing temperature, being incompatible with the temperature independence expected in a high-temperature
paramagnetic regime~\cite{AB}. The deviation from the linear behavior at high temperatures is due to the structural transition from the monoclinic to a triclinic phase at $T_\mathrm{S}=285$~K~\cite{Kohl}, giving rise to an additional relaxation channel (see
Appendix for detailed discussion).

At temperatures between 75 and 200~K, $\Delta H_{pp}(T)$ follows a critical power law,
$\Delta H_{pp}(T)\propto T^{\alpha}$ with the exponent  $\alpha= -0.81\pm 0.03$  ($-0.12\pm 0.05$) for $H\perp c$ ($H\parallel c$).  Combined with the concomitant occurrence of the weak $T-$ dependence of $H_{\mathrm{res}}(T)$, the critical-like line broadening can be related to the development of local spin correlations, representing a correlated paramagnetic state. On cooling below 70~K, the power law is changed to the quasilinear dependence $T^{0.96}$ ($T^{0.82}$) for $H\perp c$ ($H\parallel c$) while $H_{\mathrm{res}}(T)$ shifts progressively toward lower fields. Particularly noteworthy is the observation of the asymptotic $T$-linear  $\Delta H_{pp}(T)$ at the temperature interval between 25 and 75~K where the static susceptibility exhibits a rapid drop [see Fig.~1(b)]. Here, we recall that the universal $T$-linear behavior of the ESR linewidth is generic to $S=1/2$ spin chains or spin-ladder materials and pertains to one-dimensional low-energy spin excitations~\cite{Oshikawa,Ponomaryov}. Thus, the apparent $T-$ linear ESR line narrowing corroborates that spin-ladder-like correlations survive even in the presence of the three-dimensional interladder coupling~\cite{Rao}. Our ESR results are consistent with the proposed coupled two-leg ladder system, which undergoes a crossover to a decoupled ladder regime with pseudogap-type behavior at elevated temperatures~\cite{Gibbs,Rao,Troyer}. This can explain the gapped
magnetic excitations observed by a $^{125}$Te NMR study~\cite{Gibbs}.

On approaching $T_{\mathrm{N}}$,  $\Delta H_{pp}(T)$ shows a minimum at about $T_{\mathrm{min}}=25$~K and then a strong critical increase $\Delta H_{pp}(T)\propto T^{-3.0}$ ($T^{-2.2}$) for $H\perp c$ ($H\parallel c$). Finally, the ESR signal is wiped out just above $T_{\mathrm{N}}$. As the steep increase of $\Delta H_{pp}(T)$ below $T_{\mathrm{min}}$  accompanies a large upshift of $H_{\mathrm{res}}(T)$, $T_{\mathrm{min}}$ is linked to a crossover of the ladder correlations to three-dimensional spin correlations, bringing about a slowing down of spin fluctuations. In this regard, $T_{\mathrm{min}}=0.25-0.29 J$ gives
an average energy scale of the 3D interactions.

In an antiferromagnetically ordered state below $T_{\mathrm{N}}$, the anticipated antiferromagnetic resonance modes cannot be detected for the employed frequency either due to a large gap in the spin-wave excitation spectrum or to strong quantum fluctuations faster than 9.4~GHz. We are led to the conclusion that a series of power-law correlated regimes unveil the successive thermal crossover from a correlated paramagnet through a spin ladder to a 3D correlated state.

\begin{figure}
\label{figure1}
\centering
\includegraphics[width=8.5cm]{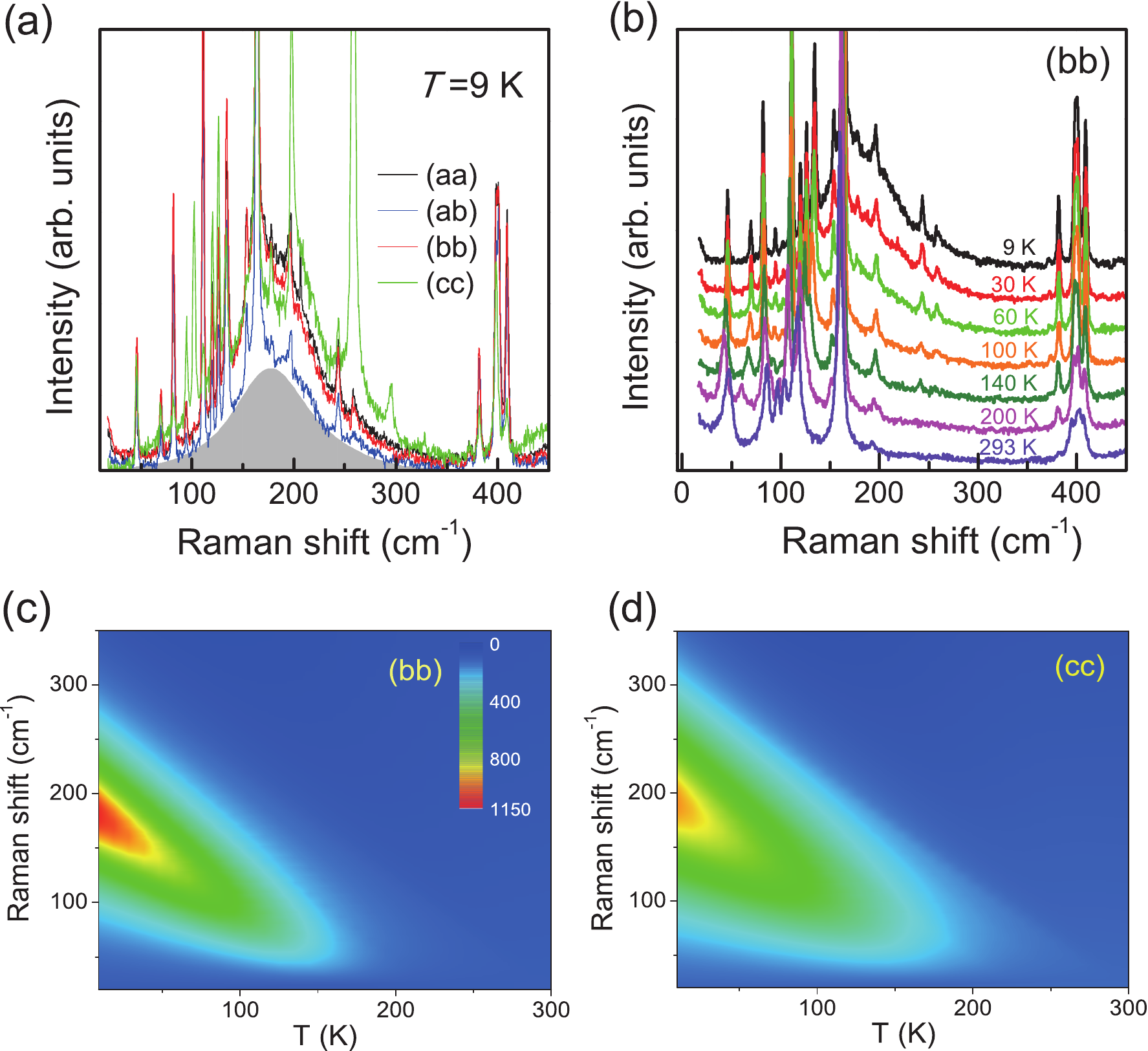}
\caption{(a) Low-energy range of Raman spectra taken at $T=9$~K in four different polarizations. The gray shading denotes a two-magnon continuum. (b) Magnetic excitations in (bb) polarization for different temperatures. The spectra are vertically shifted by a constant amount. (c),(d) Color contour plot of the magnetic Raman scattering intensity in the temperature-Raman-shift plane in (bb) and (cc) polarizations, respectively.}
\end{figure}

\subsection{Magnetic Raman scattering}
Next, we turn to the low-energy magnetic excitations probed through double spin-flip processes by inelastic light scattering.  As shown in Fig.~4(a), a two-magnon (2M) continuum (gray shading) extending from 50 to 350~cm$^{-1}$ is observed in all measured polarizations, i.e., (aa), (bb), (cc), and (ab) scattering configurations at $T=9$~K. Here, the (xx) polarization denotes the polarization of the incoming
and outgoing light parallel to the crystallographic $x$ axis. In this notation, (bb) and (cc) denote the leg and the rung polarization, respectively, and (aa) and (ab) are interladder polarizations. The magnetic spectra hardly vary in spectral form but in intensity as a function of polarization. With increasing temperature, the magnetic continuum is heavily damped and evolves to a quasielastic scattering at high temperatures as shown in Fig.~4(b) (see Appendix for the phonon modes).

Within the Fleury and Loudon theory~\cite{FL},  the $S=1/2$ two-leg ladder is predicted
to exhibit the Raman continuum with a two-peak structure in an isotropic regime~\cite{Schmidt}.
The observed single-peak continuum means that a pure isolated ladder model is not sufficient to
describe the strongly coupled spin ladder of Ba$_2$CuTeO$_6$.
As the isotropic ladders have a dominant 2D nature of local spin fluctuations, the 2M
continuum has a primary peak energy at $\omega_{\mathrm{2M}}\cong 2.7~J$ as in a case of
$S=1/2$ two-dimensional antiferromagnets~\cite{Lyons,Gozar,Gruninger,Windt}. Our peak position of $\omega_{\mathrm{2M}}=185$~cm$^{-1}$ yields the spin exchange interaction of $J/k_\mathrm{B}=98$~K. This value is comparable to $J/k_\mathrm{B}=86$~K evaluated from the nearly isotropic ladder model~\cite{Gibbs}, but is twice $J/k_\mathrm{B}=48.6$~K estimated from the modified chain model~\cite{Rao}. This estimate together the asymptotic $T$-linear  $\Delta H_{pp}(T)$ at elevated temperatures suggests that the spin-ladder model provides a first approximation to the magnetism of Ba$_2$CuTeO$_6$.

The magnetic Raman scattering intensity scales in each polarization with the strength of the exchange interactions as it is given by $I_R\propto \mid\langle i |\sum_{i,j}\mathbf{S}_i\cdot \mathbf{S}_j  \mid j\rangle|^2$. The ratio of the integrated scattering
intensity is found to be $I(aa):I(bb):I(cc):I(ab)=1:0.91:0.79:0.55$. The comparable intensity between the parallel polarizations confirms a three-dimensionally networked spin ladder with $J_r\approx J_l$~\cite{Freitas}. Shown in Figs.~4(c),(d) are the color contour plots of the magnetic Raman scattering intensity representing the temperature dependence of the Raman shift in (bb) and (cc) polarizations. The magnetic response is obtained after subtracting the sharp phonon peaks from the raw spectra. Upon heating, the 2M continuum softens and weakens in a quasilinear manner.

\begin{figure}
\label{figure1}
\centering
\includegraphics[width=9cm]{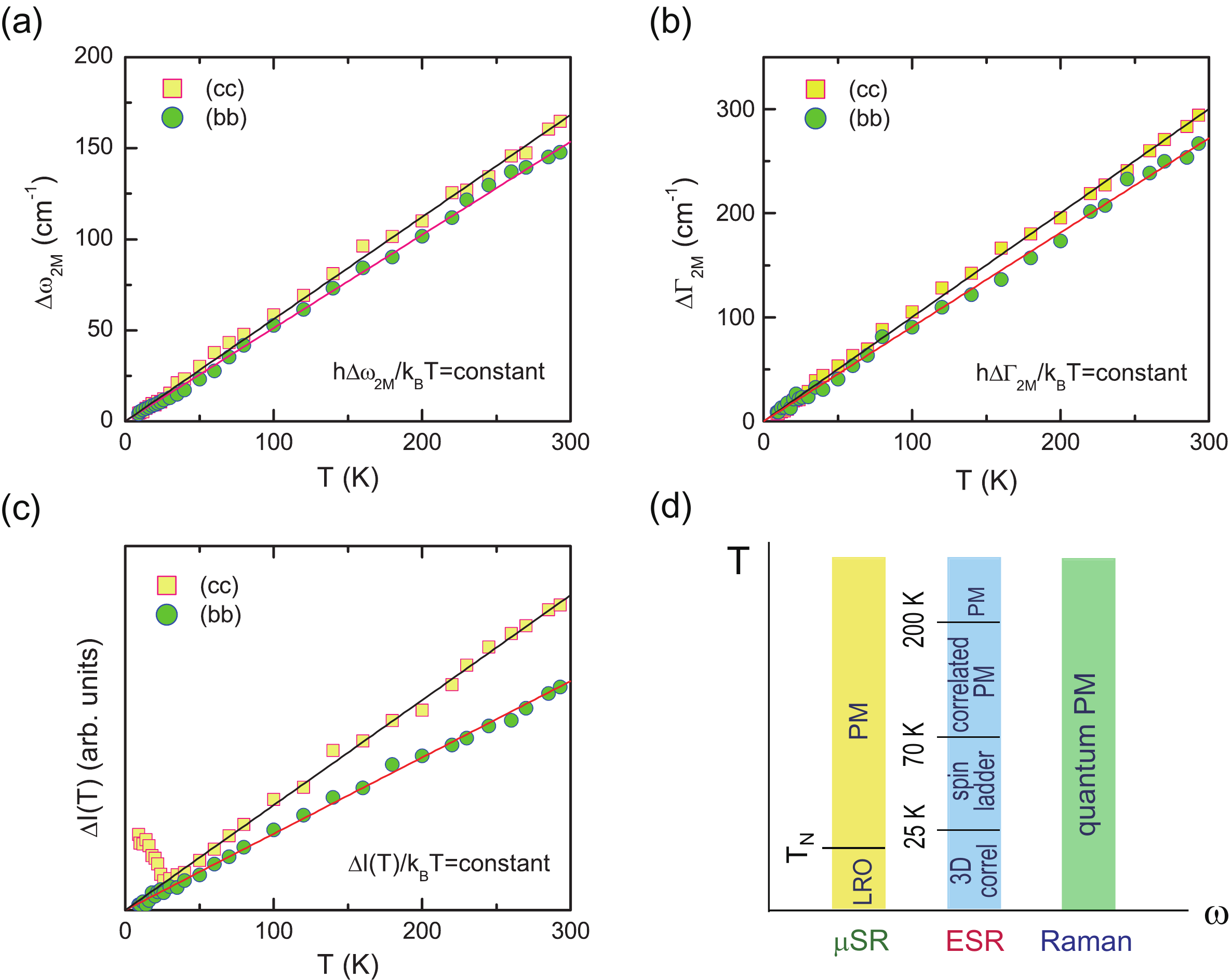}
\caption{(a),(b),(c) Temperature-dependent part of the two-magnon energy, the linewidth,
and  the integrated intensity obtained after the subtraction of zero-temperature residual terms. 
The solid lines are linear fits to the data. (d) Phase diagram
of Ba$_2$CuTeO$_6$ determined by $\mu$SR, ESR, and Raman scattering techniques in the $T-\omega$ plane.}
\end{figure}

We now detail the temperature dependence of the 2M parameters.  Generically,
the $T-$ dependence of the 2M frequency, linewidth, and intensity is given by
$\omega_{\mathrm{2M}}(T)=\omega_{\mathrm{2M},0}-\Delta\omega_{\mathrm{2M}}(T)$,
$\Gamma_{\mathrm{2M}}(T)=\Gamma_{\mathrm{2M},0}+ \Delta\Gamma_{\mathrm{2M}}(T)$,
and $I_{\mathrm{2M}}(T)=I_{\mathrm{2M},0}-\Delta I_{\mathrm{2M}}$, respectively.
The first $\omega_{\mathrm{2M},0}$, $\Gamma_{\mathrm{2M},0}$, and $I_{\mathrm{2M},0}$ terms correspond to the energy, lifetime, and intensity of the quasiparticles
at $T=0$~K. The second terms are associated with the renormalization
and damping of the quasiparticles by thermal and quantum fluctuations.
Remarkably,  as shown in Figs.~5(a)-(c), the $T-$ dependent part of the 2M parameters is determined by temperature itself: $\Delta\hbar\omega_{\mathrm{2M}}/k_BT$, $\Delta\hbar\Gamma_{\mathrm{2M}}/k_BT$, and $\Delta I_{\mathrm{2M}}/k_BT$ are constant over almost two decades of temperature (see the solid linear fits).
The observed $T-$ linear dependence means that the only relevant energy scale is 
the thermal energy $k_BT$, lacking any energy scale of the underlying microscopic description. The universal scaling of $\Delta\omega_{\mathrm{2M}}$, $\Delta\Gamma_{\mathrm{2M}}$, and $\Delta I_{\mathrm{2M}}$  suggests that the magnon excitations acquire a quantum nature at finite temperatures,
 indicating the proximity to a quantum phase transition. This is supported by our ESR data that evidence the predominance of the RVB-type correlations at elevated temperatures.

We stress that these characteristics are fundamentally different
from a thermal critical behavior of magnons and triplons. As the temperature is increased through $T_\mathrm{N}$ in conventional antiferromagnets, magnons rapidly dampen and soften by thermal fluctuations and the integrated intensity grows strongly toward saturation in a high-$T$ paramagnetic state~\cite{Cottam,Choi08}. In a case of spin gapped systems, the peak
energy and linewidth of triplons hardly vary with temperature while
their intensity is gradually suppressed due to the thermal depletion of singlets~\cite{Choi05}. In this regard,
the quasiparticles of  Ba$_2$CuTeO$_6$ are described by neither magnons nor triplons.

In Fig.~5(d), we sketch the $T-\omega$ phase diagram of Ba$_2$CuTeO$_6$. A combination of $\mu$SR, ESR, and Raman scattering techniques unveils the distinct aspect of magnetic correlations, depending on a frequency (time) window.
ZF-$\mu$SR allows identifying the static magnetic order at $T_\mathrm{N}=14.1$~K.
However, only 8 \% of the expected total
magnetic entropy is recovered across the magnetic transition at $T_\mathrm{N}$, implying that most of entropy is released above $T_\mathrm{N}$~\cite{Rao}.
The power-law dependence of the ESR linewidth $\Delta H_{pp}(T)$ with changing exponents
evidences a successive crossover from a paramagnetic through a spin-liquid-like into a 3D correlated state with decreasing temperature.
In contrast, the 2M Raman response
is hardly affected by the successive transitions
observed by $\mu$SR and ESR. Significantly, the 2M parameters obey a scaling relation
 over the entire measured temperature range.
As the 2M peak energy of $\omega_{\mathrm{2M}}\cong 2.7~J$ has a much larger energy scale than the residual 3D interactions of $0.25-0.29~J$,  the characteristic features
of the 2M response are dominated by short-time spin correlations. Thus,
the short-wavelength magnetic excitations behave like quantum quasiparticles. In a long-wavelength limit, however, thermal fluctuations become stronger, thereby the distinct magnetic correlations show up within a gigahertz frequency window.
The frequency-dependent magnetic behavior may be a generic feature of the system
which is in the vicinity of the transition to 3D antiferromagnetic order.

\section{CONCLUSIONS}
To conclude,  we have presented a combined study of ZF-$\mu$SR, ESR, and Raman scattering measurements on the coupled two-leg spin ladder Ba$_2$CuTeO$_6$. The former two resonance techniques disclose a consecutive evolution of spin correlations, suggesting the presence of several different magnetic energy scales. At elevated temperatures, a crossover to a decoupled ladder takes place. Strikingly, we find that the two-magnon Raman response representing short-time spin correlations obeys a $T$-linear scaling relation  in its parameters over a wide range of temperature. This scaling behavior signifies that the local spin correlations are governed by the competition between thermal and quantum fluctuations and that Ba$_2$CuTeO$_6$ is close to a quantum critical point from an ordered side.

\section*{ACKNOWLEDGMENTS}
This work was supported by  the Korea Research Foundation (KRF) grant funded by the Korea government (MEST) (Grant No. 20170065). 
S.H.D. was supported by Chung-Ang University Research Assistant Fellowship.

\section*{APPENDIX: Phonon Raman scattering }
\begin{figure}
\label{figure1}
\centering
\includegraphics[width=8cm]{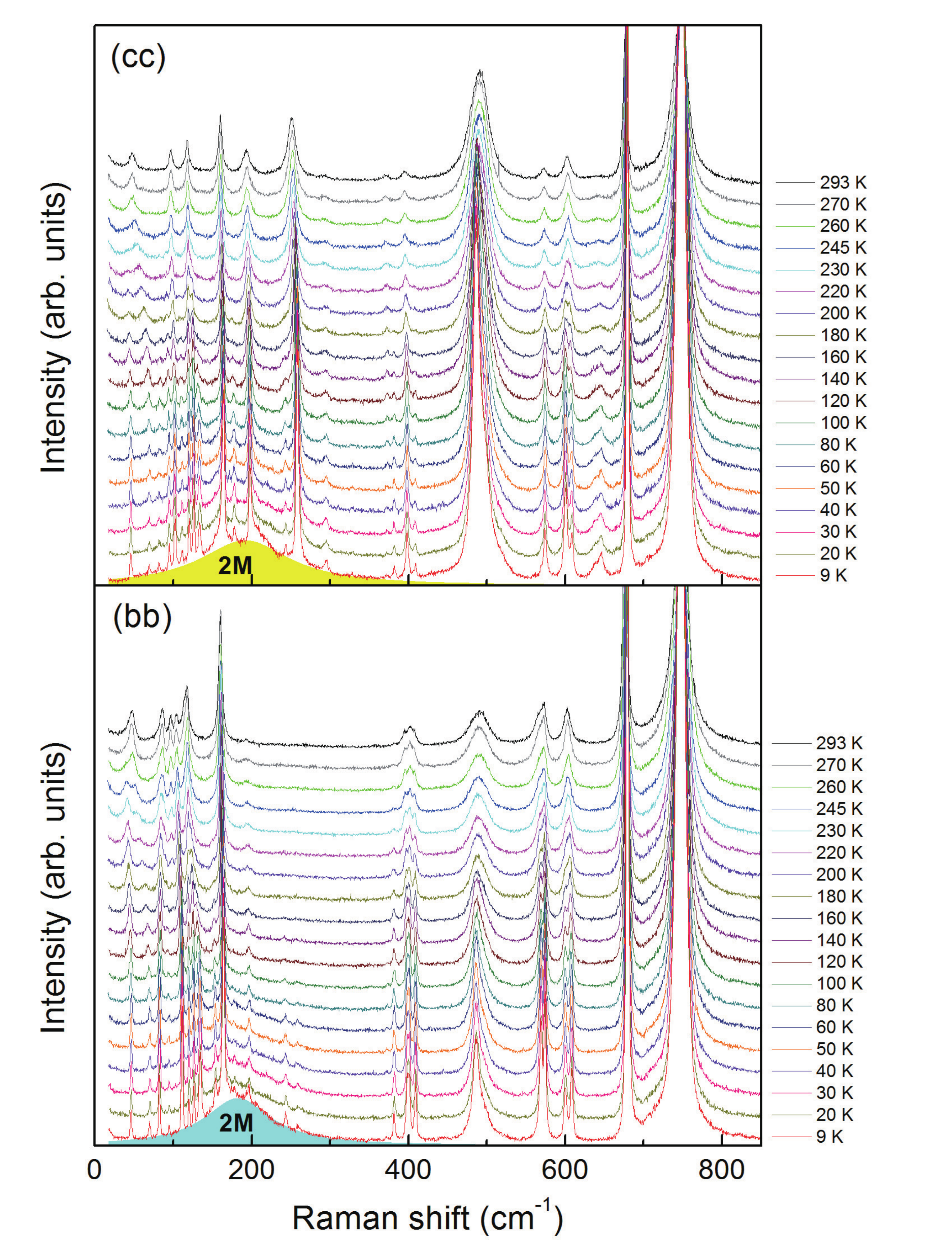}
\caption{Raman spectra measured in (cc) and (bb) polarization for different temperatures. The shadings emphasize scattering due to two-magnon excitations.}
\end{figure}

According to the structural analysis, the monoclinic $C2/m$ crystal structure of Ba$_2$CuTeO$_6$ undergoes a structural phase transition to the triclinic $\bar{P}$ structure at about 287 K~\cite{Kohl}.  The factor group analysis for the monoclinic $C2/m$ crystal symmetry yields the total irreducible representation for Raman-active modes: $\Gamma_\mathrm{R}=16A_{g} (xx, yy, zz, xz)+11B_{g} (xy, yz)$.  In the low-temperature triclinic phase, one expects the total of 27 one-phonon Raman-active modes: $\Gamma_\mathrm{R}=27A_{g}(xx, yy, zz, xy, xz, yz)$.

Figure~6 presents the temperature dependence of the Raman spectra of Ba$_2$CuTeO$_6$ measured in (cc) and (bb) polarizations in the frequency range of $17-850\, \mbox{cm}^{-1}$. At room temperature, we observe $15A_{g}$ phonon modes and in the low-temperature phase $27A_{g}$ phonon peaks, which agree well with the factor group predictions. This confirms the occurrence of the structural phase transition. The sharp phonons are superimposed on top of a broad continuum centered at 185 cm$^{-1}$ (denoted by shadings). The continuum is ascribed to a two-magnon excitation, judging from its energy scale and temperature dependence. With increasing temperature, the two-magnon continuum shifts to lower energies and is systematically suppressed. We note that the peak energy of the two-magnon scattering is larger than twice the spin gap of $2\Delta=92$~K, advocating the presence of sizable inter-leg interactions beyond a simple isolated two-leg model. Further discussions on the two-magnon scattering are provided in the main text.

\begin{figure*}
\label{figure1}
\centering
\includegraphics[width=14cm]{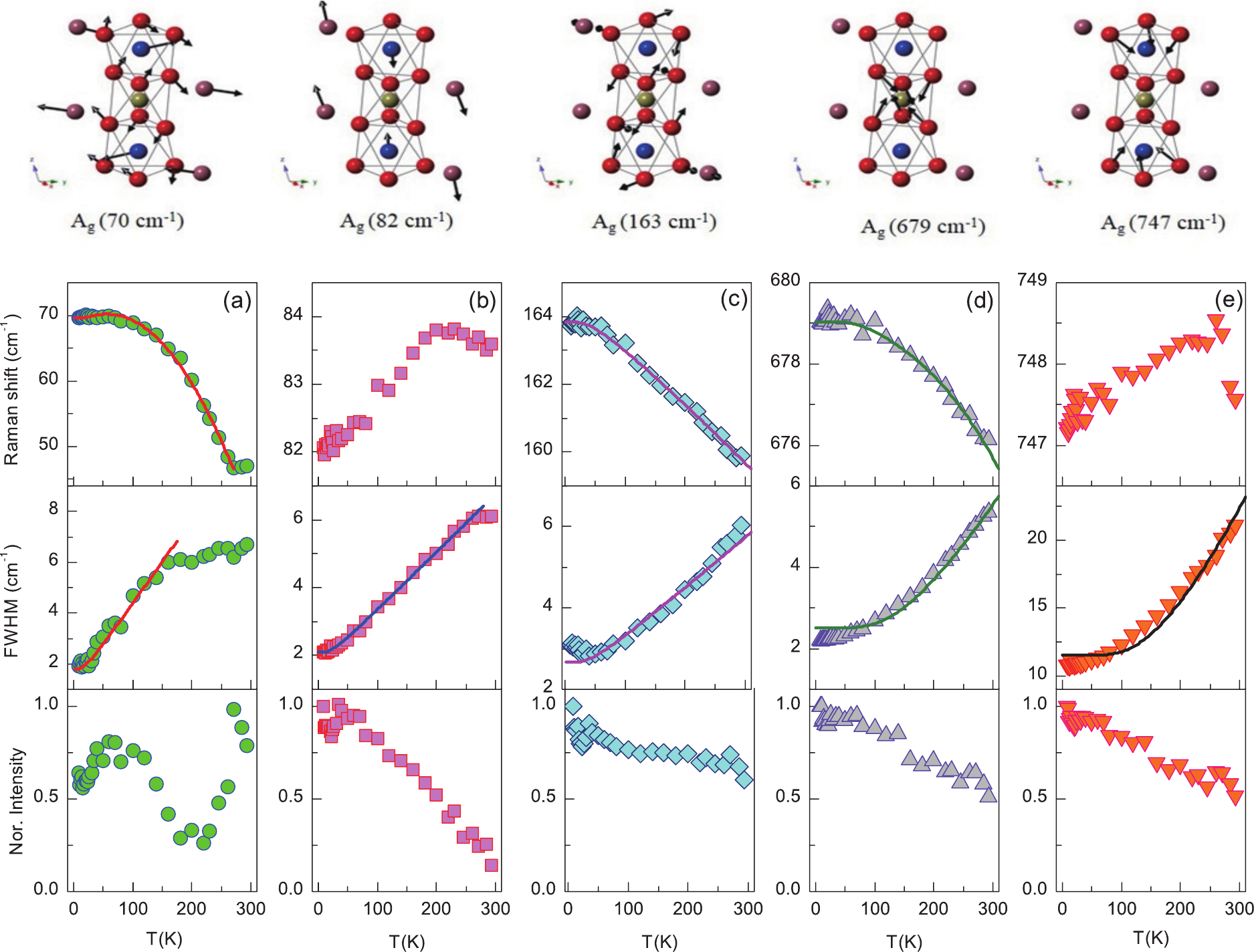}
\caption{ Temperature dependence of the frequency, the linewidth, and the normalized intensity of the (a) 69.6, (b) 82.1, (c) 163.8, (d) 679 and (e) 747.2 cm$^{-1}$ modes taken in (bb) polarization. The upper panel presents a schematic representation of the Raman-active normal modes for the respective phonons. The relative amplitudes of the vibrations are represented by the arrow lengths. The blue, green, purple, and red balls stand for Cu, Te, Ba, and O atoms, respectively.}
\end{figure*}

\begin{figure}
\label{figure1}
\centering
\includegraphics[width=8cm]{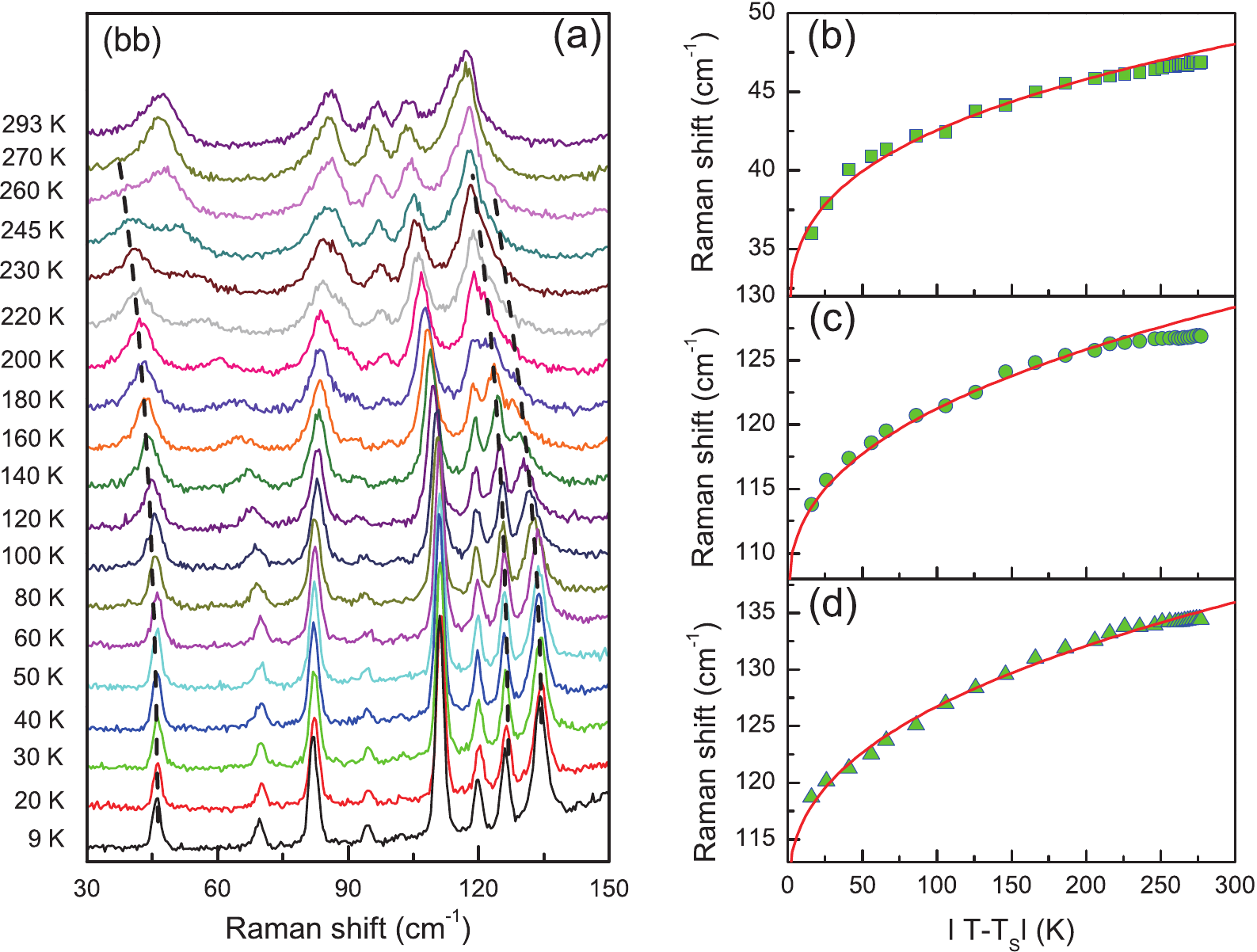}
\caption{(a) Soft modes in ($bb$) polarization as a function of temperature. Temperature dependence of the frequencies of the soft modes at (b) 47, (c) 126, and (c) 134 cm$^{-1}$. The solid lines are a fit described in the text.}
\end{figure}
Next, we pay our attention to the phonon peaks. The phonon modes can be classified into three spectral regimes in accordance with the frequency separation: (I) $40-300\, \mbox{cm}^{-1}$, (II) $350-500\, \mbox{cm}^{-1}$, and (III) $550-800\, \mbox{cm}^{-1}$. The low-energy phonon bands involve the displacement of the heaviest Ba atoms and the rotations of the CuO$_{6}$ and the TeO$_{6}$ octahedra. The intermediate-energy bands are assigned to the bending vibrations of the CuO$_{6}$ and the TeO$_{6}$ octahedra. The high-energy bands correspond to the breathing vibrations of the CuO$_{6}$ and the TeO$_{6}$ octahedra. Upon heating, the phonon modes exhibit intriguing changes in number, frequency, and linewidth as a function of temperature. To quantify their evolution as a function of temperature, we fit them to a sum of the Lorentzian profiles. The temperature dependence of the resulting phonon parameters is summarized in Fig.~7 for the five representative peaks at (a) 69.6, (b) 82.1, (c) 163.8, (d) 679 and (e) 747.2 cm$^{-1}$ together with their normal modes.

Upon heating, the 69.6 cm$^{-1}$ mode exhibits a giant softening by about 22.5 cm$^{-1}$, being in stark contrast to the weak temperature dependence observed in other modes. The observed frequency shift is much larger than the several cm$^{-1}$ expected from lattice anharmonicity. As this mode involves the combined rotational vibrations of the CuO$_{6}$ octahedra about the $x$ axis and out-of-phase motions of the Ba and Cu atoms in the $xz$ plane (see Fig. 7), the large softening implies that the structural phase transition brings about the substantial octahedral tilting and rotations. The temperature dependence of the linewidth cannot be described by an anharmonic model over a larger temperature range (see the solid line)~\cite{Balkanski}. In addition, the normalized intensity shows a nonmonotonic variation with temperature, indicating a modulation of the electronic polarizability induced by the structural phase transition. The 82 cm$^{-1}$ mode involves out-of phase vibrations of the Ba and Cu atoms in the $yz$ plane. Remarkably, with increasing temperature the 82 cm$^{-1}$ mode displays even a hardening by 1.7 cm$^{-1}$ with the subsequent small softening by 0.5 cm$^{-1}$ above 180 K. This behavior is opposite to the anticipated hardening. The small anomaly at about 35 K is observed in the peak position, the linewidth and the normalized intensity.

The 163 cm$^{-1}$ mode is associated with twisting vibrations of the CuO$_{6}$ octahedra and out-of phase motions of the Ba(2) atoms along the $x$ axis. The 679 cm$^{-1}$ mode corresponds to bending vibrations of the TeO$_6$ octahedra. For both modes, the phonon parameters are largely described by the anharmonic model. Small anomalies in the peak position, the linewidth and the normalized intensity are discernible at about 35 K. The 747 cm$^{-1}$ mode involving stretching vibrations of the threefold octahedral blocks displays a similar temperature dependence found in the 82 cm$^{-1}$ mode. Overall, the phonon modes show an intriguing temperature dependence reflecting the structural transition.

The direct consequence of the second-order phase transition is the presence of soft modes in the A$_g$ symmetry.  Indeed, we are able to identify at least three soft modes at 47, 126 and 134 cm$^{-1}$ as plotted in Fig.~8(a). Upon approaching the phase transition, these phonons shift to lower energies and finally disappear at about $T_{\mathrm{S}}=285$~K [see the dashed lines in Fig.~8(a)]. Their frequencies are plotted in Figs.~8(b)-(d) as a function of $|T-T_\mathrm{S}|$. Fitting these energies to a power law, $\omega(T)=\omega(0~ \mathrm{K})+|T-T_\mathrm{S}|^{\beta}$, gives a reasonable description with $\beta=1/3\pm 0.02$ [see the solid lines in Figs.~8(b)-(d)]. This is far from a mean-field value of $\beta=1/2$.

%\bibliography{achemso}

\end{document}